# Experimental demonstration of an anomalous Floquet topological insulator based on negative-index media

ZHIWEI GUO,[1] YONG SUN,[1,3] HAITAO JIANG,[1] YA-QIONG DING,[2] YUNHUI LI,[1] YEWEN ZHANG,[1] AND HONG CHEN[1,4]

[1]*MOE Key Laboratory of Advanced Micro-Structured Materials, School of Physics Science and Engineering, Tongji University, Shanghai 200092, China*
[2]*Science College, University of Shanghai for Science and Technology, Shanghai 200093, China*
[3]*e-mail: yongsun@tongji.edu.cn*
[4]*e-mail: hongchen@tongji.edu.cn*

**Photonic and acoustic topological insulators exhibiting one-way transportation that is robust against defects and impurities are typically realized in coupled arrays of two-dimensional ring resonators. These systems have produced a series of applications, including optical isolators, delay lines, and lasers. However, the structures are complicated because an additional coupler ring between neighboring rings is needed to construct photonic pseudo-spin. In this work, a photonic analogue of the quantum spin Hall effect is proposed and experimentally demonstrated in an anomalous Floquet topological insulator in the microwave regime. This improved design takes advantage of the efficient and backward coupling of negative-index media. The results contribute to the understanding of topological structures in metamaterials and point toward a new direction for constructing useful topological photonic devices.**

© 2020 Chinese Laser Press

http://dx.doi.org/10.1364/PRJ. XX. XXXXXX

## 1. INTRODUCTION

Edge modes of photonic topological insulators, which travel along the boundary of the structure and are robust against perturbations, have recently become a subject of much interest [1-3]. Although early designs for photonic topological one-way edge modes involved breaking the time-reversal (T-reversal) symmetry in gyromagnetic materials using an external magnetic field [4-7], it was soon discovered that a synthetic magnetic field could exhibit the same effect as a real magnetic field in producing one-way edge modes [8-10]. Much like the quantum spin Hall effect (QSHE) in electric systems [11, 12], T-reversal symmetric preserved photonic systems with nontrivial topological phases have attracted considerable interest from researchers. Such systems have been realized in a variety of structures, including bi-anisotropic metamaterials [13-17], photonic crystals [18-23], and arrays of coupled ring resonators (CRRs) [24-36]. Great progress in topological photonics has extended its scope from microwave to X-ray frequencies [37], passive to active structures [38, 39], near-field to far-field measurement [40, 41], short-range to long-range coupling [42-44], one to higher dimensions [45-47], and classical to quantum optics [48-50].

Diversified CRR arrays, which provide a suitable platform for the study of topological photonics, have received substantial attention. The roles of pseudo-spins in opposite directions are played by the clockwise and anti-clockwise propagation directions of light [24-32]. High-performance topological optical devices based on CRR arrays, such as delay lines [24], optical isolators [28], and lasers [31, 32], have been proposed as key components for optical communication systems. In 2013, Liang and Chong demonstrated theoretically that a lattice of optical ring resonators can exhibit a topological insulator phase even if all bands have zero Chern number [26]. Unlike earlier designs that required delicately tuning the aperiodic couplings of the inter-resonators [24, 25], this simplified structure is periodic and contains identical coupled ring resonators [26, 27]. Inspired by this anomalous Floquet topological insulator, topological edge states with *T*-symmetry have been experimentally measured using a surface plasmon structure [30]. This novel topological edge mode also has been demonstrated in acoustic systems [51, 52]. However, the coupling between the site rings in CRR arrays needs to be very strong for the realization of an anomalous Floquet topological insulator [26]. At present, the typical solution is to insert a coupler ring between two site rings. Achieving strong coupling relies on the resonance of the coupler ring, which severely limits the bandwidth of the strong-coupling regime [29]. In addition, the CRR array design is complicated by the requirement of an additional coupler ring between nearest neighboring rings in order to construct the photonic pseudo-spin [24-36]. Thus, a question naturally arises: can an anomalous Floquet topological insulator be constructed without additional coupling rings and with a broad bandwidth?

Metamaterials, artificial materials composed of subwavelength unit cells, provide a powerful platform for manipulating the propagation of light. The backward and efficient coupling mechanism of negative-index (left-handed) media can be used to solve the above problems [53-56]. Recently, high-performance metamaterials have been constructed to achieve the required optical response in transmission line (TL) platforms and have enabled extensive applications, such as cloaking [57], hyperbolic dispersion [58, 59], quantum-optics-like phenomena [60, 61], and topological photonics [21, 62]. Utilizing a TL system, in this work we design circuit-based negative-index media in the microwave regime. An improved anomalous Floquet topological insulator is constructed from a square array of CRRs composed of composite right/left-handed (CRLH) TLs. In each individual ring resonator, the roles of opposite pseudo-spins are played by the clockwise and anti-clockwise propagation of electromagnetic waves. A directional coupler inserted into the site rings supports backward wave coupling. Therefore, the direction of energy flow in each ring is same for a given pseudo-spin state, without requiring additional coupling rings. A photonic analogue of the QSHE is observed in both numerical simulations and experiments. We also investigate the robustness of edge states against perturbations in the structure. Our findings not only provide a useful understanding for studying topological structures using metamaterials, but also may offer a new means of realizing topological photonic devices.

## 2. DESIGN AND METHOD

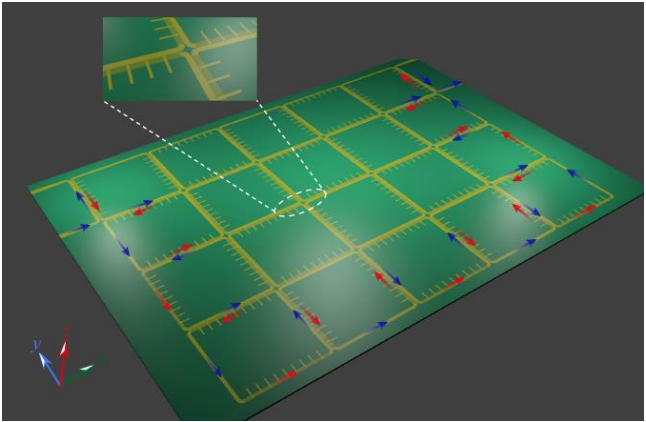

Fig. 1. The schematic of anomalous Floquet topological insulator based on CRLH ring resonators. The size of our fabricated sample is $4\times 5$ unit cells. The right-handed material (simple line strips) and left-handed material (complex branched strips) connect successively in a ring structure. The zoom in picture of different ring resonators is presented on the top. The arrows represent the coupling direction of the energy flow in two materials with different handedness. The upper metallic microstrip and the lower dielectric substrate are marked by yellow and green, respectively.

A schematic of a 4 × 5 array of CRLH ring resonators forming an anomalous Floquet topological insulator is shown in Fig. 1. Distributed composite CRLH TLs [54, 55] are used to design a left-handed media with a negative refraction index. The ring resonator can then be constructed by connecting left- and right-handed TLs in sequence. A square array of coupled CRLH ring resonators can be produced by arranging the resonators in close proximity. The size of the fabricated sample is 340.4 mm × 272.3 mm. Unlike previous anomalous Floquet topological insulators, in which additional resonators were adopted as coupling rings [24-36], every ring resonator in the improved platform belongs to a site ring, as shown in Fig. 1. Two U-shaped waveguides serve as input and output ports. In the improved platform, the site rings are backward coupled with their neighbors through the vortex-like interface mode existing in the CRLH coupler, shown in Fig. 2(a). This provides considerable advantages, such as high coupling strength and broad bandwidth [53-56]. It is the mechanism of backward coupling that lead to the electromagnetic waves will not coupling return to the input ring after the coupling process once, which greatly increases the coupling efficiency [53-56]. Additional coupler rings are not required in anomalous Floquet topological insulators based on left-handed metamaterials. Details of the structural parameters, effective circuit models, and the design scheme of the negative-index media are given in the Appendix A.

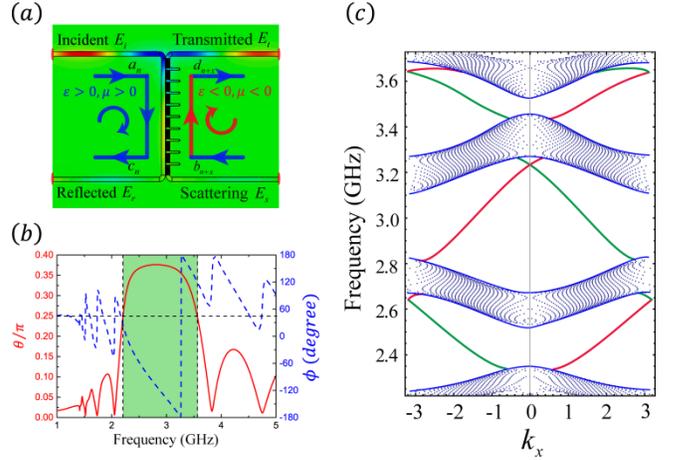

Fig. 2. (a) Configuration of U-U waveguide structure. The incident wave with high performance is coupled to the transmitted port based on the wave vector conservation condition for back direction. (b) The value of coupling strength $\theta$ (red line) and the corresponding coupling phase (blue dashed line) can be derived from the U-U shape structure. (c) The projected band diagram of a semi-infinite strip coupled ring lattice, whose width is finite in the y direction ($N$=30) and periodic in the $x$ direction. Gapless edge modes periodically appear in the band-gap range of the infinite periodic structure. Red and green lines denote edge states corresponding to the upper and lower edges of the sample, respectively.

The band structure of the proposed array of CRLH ring resonators can be calculated as follows. The amplitudes of the waves in the ring resonators can be expressed in the form of a scattering matrix [26]:

$$S(\theta,\chi) = e^{i\chi}\begin{bmatrix} \sin\theta & i\cos\theta \\ i\cos\theta & \sin\theta \end{bmatrix}, \quad (1)$$

where $\theta$ denotes the coupling strength between adjacent lattice rings and $\chi$ is the overall phase factor. The highly dispersive behavior of $\theta$, plotted in Fig. 2(b), can be written as:

$$\theta = \sin^{-1}[(I_{out}/I_{in})^{1/2}], \quad (2)$$

which is derived from the U-U shape structure in Fig. 2(a) using the CST (computer simulation technology) Microwave Studio software. $I_{in}$ and $I_{out}$ denote the power thorough the incident and transmitted waveguides, respectively. The green region marked in Fig. 2(b) covers the frequency range for strong coupling, where $\theta > \pi/4$; the nontrivial topological phase exists in this range [26-30]. Edge states can be observed in a semi-infinite strip coupled rings lattice, whose width is finite in the $y$ direction and periodic in the $x$ direction. The positions of lattice rings are indicated by $n=(x_n, y_n)$, where $x_n$ and $y_n$ are the column and row indices, respectively. The coupling between the lattice rings at sites $n$ and $n+x$ ($n+y$) in the $x$ ($y$) direction is expressed by the scattering matrixes $S_x$ ($S_y$):

$$S_x \begin{bmatrix} a_n \\ b_{n+x} \end{bmatrix} = \begin{bmatrix} d_{n+x} \\ c_n \end{bmatrix}$$
$$S_y \begin{bmatrix} d_n \\ c_{n+y} \end{bmatrix} = \begin{bmatrix} b_{n+y} \\ a_n \end{bmatrix} e^{-2i\phi}. \quad (3)$$

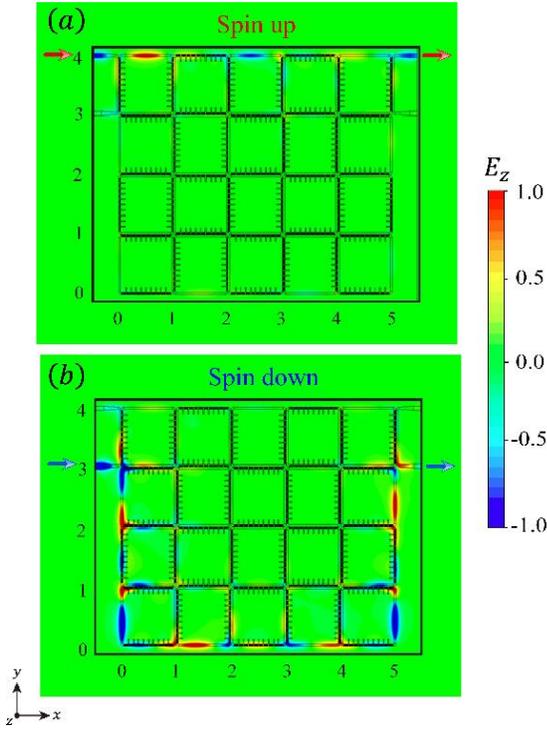

Fig. 3. The simulated electric field $E_z$ distributions of the topological edge states at 2.5 GHz. (a) The simulated $E_z$ distribution when pseudo-spin-up edge state is selectively excited and the electromagnetic wave transports along the upper edge of the structure. (b) Similar to (a), but for the one-way pseudo-spin-down edge state. In this case, the electromagnetic wave transports along the lower edge of the structure.

The Bloch condition can be applied in the $x$ direction for the periodic condition $|\varphi_{n+x}\rangle = e^{iK_x}|\varphi_n\rangle$. Combining this with the boundary conditions in the $y$ direction, $c_1 = e^{-2i\phi}b_1$ and $a_N = e^{2i\phi}d_N$, the equation for the eigenvalues is obtained [26]:

$$\begin{bmatrix} M_1^x & & & \\ & M_2^x & & \\ & & \ddots & \\ & & & M_N^x \end{bmatrix} \cdot \begin{bmatrix} e^{-2i\phi} & & & \\ & M_1^y & & \\ & & \ddots & \\ & & & M_{N-1}^y \\ & & & & e^{2i\phi} \end{bmatrix} \cdot \psi = e^{iK_x}|\Psi\rangle. \quad (4)$$

where $|\psi\rangle = [b_1 \ d_1 \cdots b_N \ d_N]^T$ and the sub-matrices are

$$M_i^x = \begin{bmatrix} \csc\theta & -i\cot\theta \\ i\cot\theta & \csc\theta \end{bmatrix} \quad (i=1,2,\cdots N)$$
$$M_j^y = \begin{bmatrix} ie^{2i\phi}\sec\theta & -i\tan\theta \\ i\tan\theta & -ie^{-2i\phi}\sec\theta \end{bmatrix} \quad (j=1,2,\cdots N-1). \quad (5)$$

According to Eqs. (4) and (5), gapless edge modes appear in the band-gap range, as plotted in Fig. 2(c). The red and green lines denote edge states corresponding to the upper and lower edges. The emergence of a topological edge state is one of the most striking properties of topological insulators. We next demonstrate the photonic analogue of the QSHE in the improved anomalous Floquet topological insulator. Each individual ring resonator possesses two-fold degenerate modes, corresponding to a two-fold pseudo-spin degree of freedom in which the clockwise and anti-clockwise propagation directions of light play the role of the opposite pseudo-spins. By choosing the appropriate input port, clockwise and anti-clockwise circulating photonic modes (e.g., two types of pseudo-spin) can be selected. As a result, two one-way edge modes can be excited along opposite directions. Figure 3 show the results of full wave simulations of the electric field ($E_z$) distributions of the edge states. When the upper arm of the incident U-shaped waveguide (marked by the red arrow in Fig. 3(a)) is selected as the input port, the spin-up edge mode is excited with a working frequency of 2.5 GHz. Owing to the spin-orbit coupling of the QSHE, the wave is transmitted to the output port only along the upper portion of the array structure. The electric field distribution of the $E_z$ component is displayed in Fig. 3(a). Similarly, when the lower arm of the incident U-shaped waveguide (marked by the blue arrow in Fig. 3(b)) is chosen as the input port, the spin-down edge mode is selectively excited and the wave is transmitted only along the lower portion of the array structure, as shown in Fig. 3(b). More details regarding the robustness of the topological edge states are discussed in the Appendix B. Our results provide a way to design the novel topological photonic devices with excellent performance by using the negative-index media.

The experimental sample based on TLs is constructed according to the scheme in Fig. 1. The experimental setup is shown in Fig. 4. The signals are generated by a vector network analyzer (Agilent PNA Network Analyzer N5222A) and then input to the sample, which functions as the source for the system. The normalized transmission spectra for the pseudo-spin-up (red line) and pseudo-spin-down (blue dashed line) cases are shown in Fig. 5. The broadband nature of the topological edge states is indicated by the three bands with high transmittance, consistent with the dispersion relation in Fig. 2(c). In our experiments, the samples are placed on an automatic translation device, allowing the field distribution to be accurately probed using a near-field scanning measurement. The signal is generated from input of a vector network analyzer. A monopole antenna connected to output of the analyzer serves as a near-field probe to record the electric field distribution. The antenna has a length of 1 mm and is placed vertically 1 mm above the TLs to measure the signal from the electric fields of the TLs in the 2D plane. The spatial distribution of the near field is scanned in steps of 1 mm in both the $x$ and $y$ directions. The field amplitudes are normalized according to their respective maximum amplitudes.

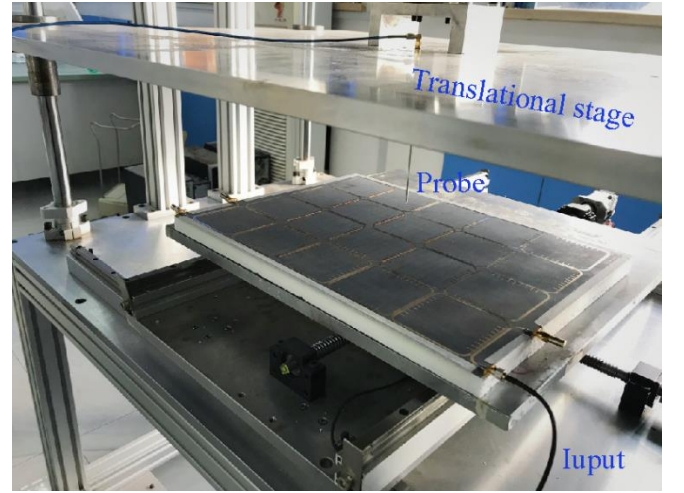

Fig. 4. Photo and diagram of experimental setup for the measurement of the vertical electric field distributions. Our experimental setup is composed of a vector network analyzer, a 2-D translational stage, and the sample to be measured. The sample is put on 2-cm-thick foam substrate with a permittivity of near one. The electric probe is a home-made rod antenna, which is connected to the output port of the vector network analyzer.

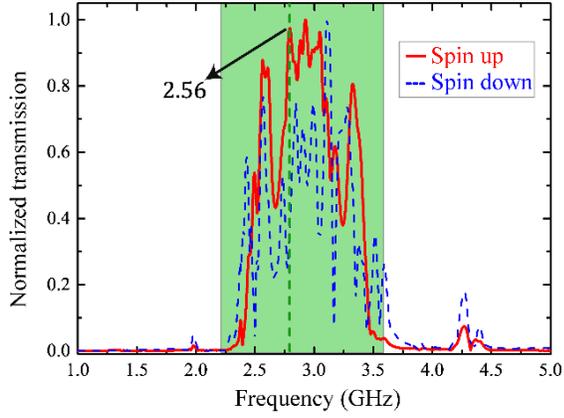

Fig. 5. Measured normalized transmission spectra of pseudo-spin-up (red line) and pseudo-spin-down (blue dotted line) cases based on the configuration in Fig.4.

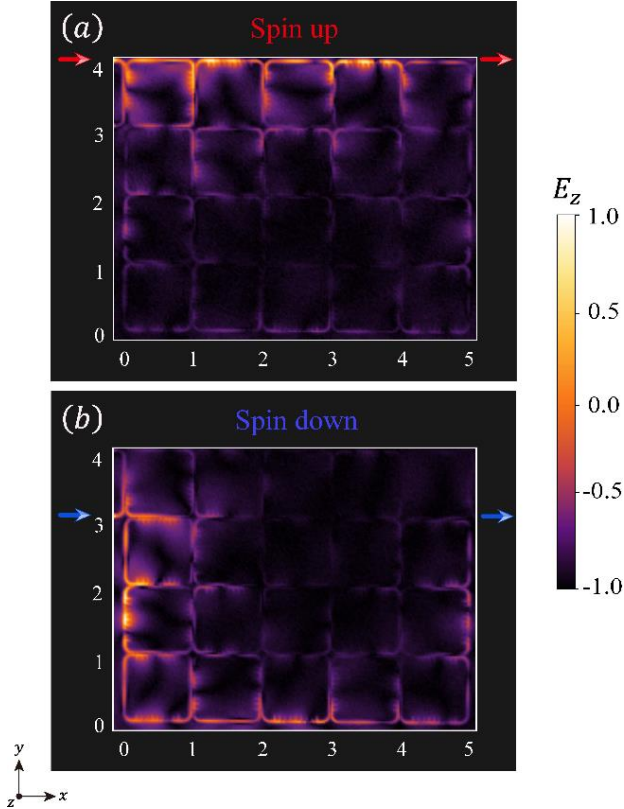

Fig. 6. Measured 2D vertical electric field $E_z$ distributions of the one-way edge states at 2.56 GHz. The exciting position of the pseudo-spin-up (pseudo-spin-down) edge state is marked by the red (blue) arrow.

A frequency of 2.56 GHz is chosen, marked by the green dashed line in Fig. 5. The edge state with pseudo-spin-up propagates along the upper edge while the edge state with pseudo-spin-down propagates along the lower edge, as shown in Fig. 6(a) and Fig. 6(b), respectively. The observed decay of the topological edge states mainly arises from the absorption in the CRLH TLs. Our results not only provide a useful understanding for studying topological structures using metamaterials, but also may pave the way toward realizing broadband topological photonic devices with topological protection in the microwave regime. Furthermore, by using the mechanisms of multiple scattering in single negative metamaterials [62, 63] or the effective negative coupling coefficient [64, 65], our design also can be extended to higher frequencies and acoustical systems.

## 3. CONCLUSIONS

In summary, using circuit-based double-negative media in a TL system, we propose an improved platform to produce an anomalous Floquet topological insulator in a square array of CRLH ring resonators. Additional coupler rings are not necessary and the resulting topological edge state is broadband in nature. The robustness of topological edge states against a variety of defects has also been verified. Our experimental demonstration of selectively excited one-way edge states will further enrich the design of anomalous Floquet topological insulators and may pave the way toward designing novel photonic topological devices. Moreover, our methods have the potential to explore the topological properties in non-Hermitian systems [66], which may be useful to some applications, such as beam splitter [67], funneling of energy [68], and so on.

## APPENDIX A: THE STRUCTURAL PARAMETERS OF CRLH RING RESONATORS

The 2D TLs are fabricated on a PTFE ($\varepsilon_r = 2.2$, $\tan\theta = 0.009$) substrate with a thickness of $h = 1.57$ mm. A schematic diagram is shown in Fig. 7. The geometric parameters are $p = 6.2$ mm, $l_c = 5$ mm, $c = 2.4$ mm, $l_s = 8$ mm, $W_s = 8$ mm, and $a = 2.25$ mm. In addition, the metallic linewidth of the right-handed material is $w = 2.4$ mm. The coupling distance between two adjacent resonators is $s = 0.1$ mm and the diameter of the metal through hole (marked by the red dots in Fig. 7) is $d = 0.4$ mm.

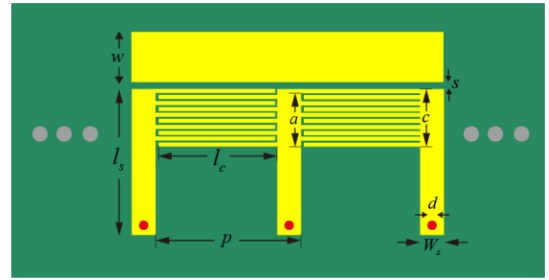

Fig.7. Structural parameters of the right-handed material (simple strip on the upper part) and left-handed material (complex branched strip on the lower part). The metallic line width of the right-hand material is $w = 2.4$ mm. The coupling distance between two different rings resonators is $s = 0.1$ mm. The geometric parameters of the left-handed material are $p = 6.2$ mm, $l_c = 5$ mm, $c = 2.4$ mm, $l_s = 8$ mm, $W_s = 8$ mm and $a = 2.25$ mm. In addition, the diameter of the metal through hole (marked by red dots) of the left-handed material is $d = 0.4$ mm.

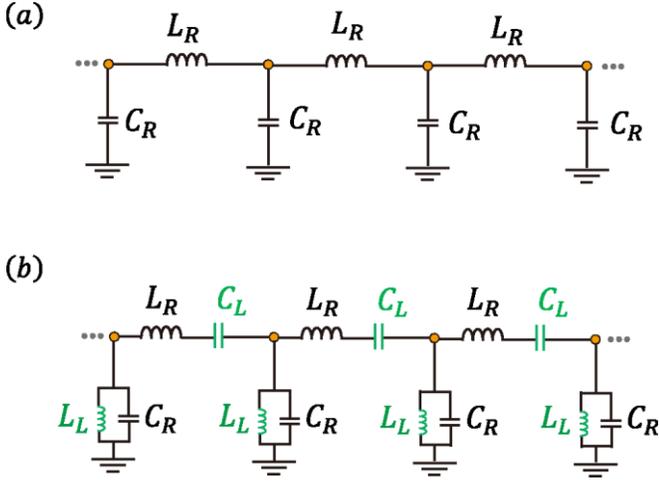

Fig. 8. (a) For the right-handed material, the effective parameters of the series inductors and shunt capacitors are $L_R \approx 2.45$ nH and $C_R \approx 0.50$ pF, respectively. (b) For the effective left-handed material, in addition to the series inductors and shunt capacitors, the series capacitors and shunt inductors are $C_L \approx 0.68$ pF and $L_L \approx 3.38$ nH, respectively.

Effective circuit models for the right- and left-handed materials are given in Fig. 8(a) and 8(b), respectively. The extracted left- and right-handed parameters are $L_L \approx 3.38$ nH, $C_L \approx 0.68$ pF, $L_R \approx 2.45$ nH, and $C_R \approx 0.50$ pF [53-56]. In the circuit, the structure factor is

$$g = [1.393 + w/h + 0.667\ln(w/h) + 1.444]^{-1}. \quad (6)$$

In the long-wavelength limit, the effective permittivity and permeability of the TLs can be written as [53-56]:

$$\begin{aligned}\varepsilon &= (C_0 - 1/\omega^2 \cdot L_L)g/\varepsilon_0 \\ \mu &= (L_0 - 1/\omega^2 \cdot C_L)/g/\varepsilon_0\end{aligned} \quad (7)$$

where $\varepsilon_0$ and $\mu_0$ are the permittivity and permeability of vacuum. $L_0$ is the inductance per unit length and $C_0$ is the capacitance per unit length of the TL segment. The negative refractive index of the left-handed material is displayed in Fig. 9(a). To study the mechanism of backward wave-vector coupling, the dispersion relationships of the left- and right-handed materials are indicated by the red solid and green dashed lines in Fig. 9(b).

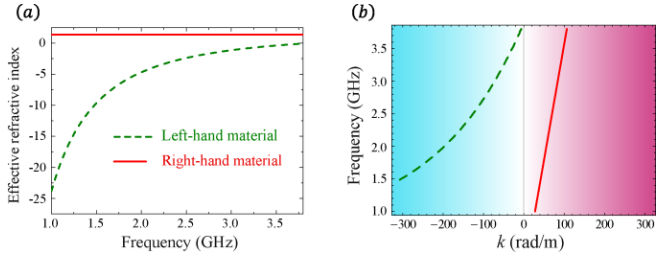

Fig. 9. (a) Effective refractive index of right-handed material (red line) and left-handed material (green dashed line). (b) Dispersion relationships of two kinds of materials in (a). The signs of the wave-vector in two kinds of materials are opposite.

## APPENDIX B: ROBUST TRANSPORTATION OF THE ONE-WAY EDGE STATES AND THE INTERNALLY EXCITED LOCALIZED STATES

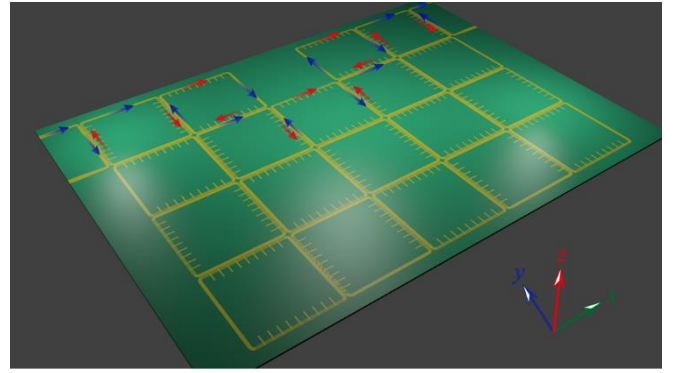

Fig. 10. The schematic of robust property of edge state for the structure perturbation when a ring resonator is removed from the upper edge of the sample. For the selectively excited one-way spin-up edge state, the wave can flow around the defect along the propagation path and continue to propagate without any distortion because of the topological protection.

One of the most important properties of photonic topological insulators is the robust one-way propagation of the edge states. In order to verify the robustness of the unidirectional edge state, we design an imperfect structure in which a ring resonator is removed from the upper edge, as presented in Fig. 10. The corresponding simulated electric field distributions of the spin-up and spin-down edge states are shown in Figs. 11(a) and 11(b), respectively. These simulations clearly show that the selectively excited one-way topological edge state flows around this special defect and continues to spread without distortion.

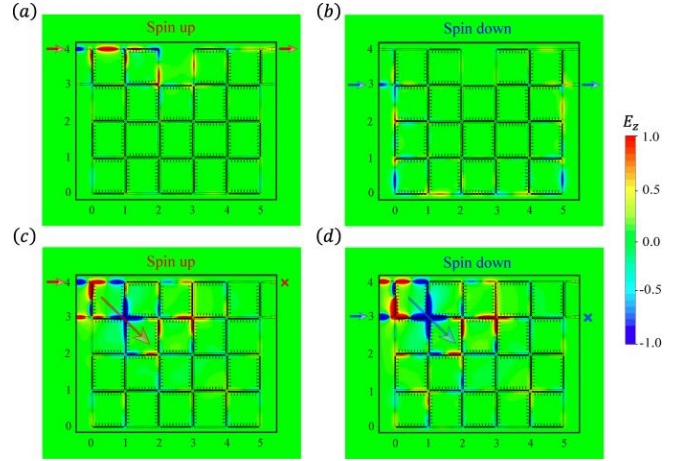

Fig. 11. (a, b) The simulated electric field distribution $E_z$ for the edge modes with topological protection (working frequency: 2.5 GHz) when the defect is achieved by fully removing a resonator ring in the upper edge. All of the selectively excited spin-up and spin-down edge modes are insensitive to the defect existing on the upward edge. (c, d) Similar to (a, b), but the working frequency is 3.1 GHz which belongs to the bulk mode.

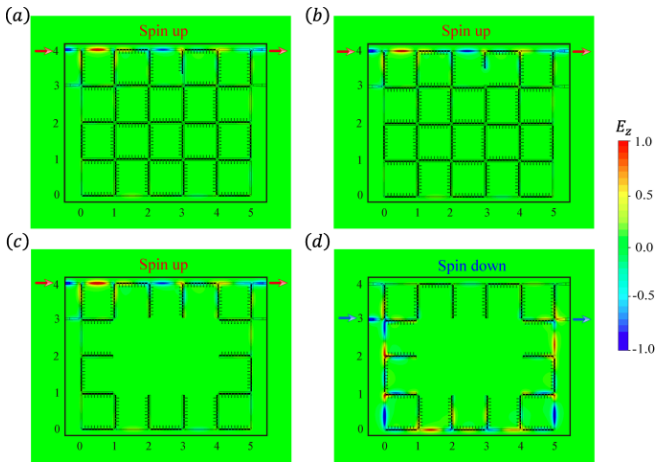

Fig. 12. (a) The selectively excited pseudo-spin-up topological edge state is maintained when a part of left-handed material is cut off on the upper edge. (b) Similar to (a), but both the left-handed and right-handed materials have been cut off. In this situation, the topological edge state also has not been destroyed. (c, d) As a special defect structure, the interior of the structure is hollowed out. For the selectively excited one-way pseudo-spin-up (spin-down) topological edge states, the wave can propagate on the upper (lower) edge as in the case without defects.

In contrast, the pseudo-spin-dependent one-way transportation behavior does not occur for the bulk modes in the pass band. As seen in Figs. 11(c) and 11(d), for a frequency in the pass band, e.g., $f = 3.1$ GHz, the incident wave flows into the bulk. There is no obvious difference between exciting the upper or the lower part of the structure.

To further demonstrate this property in addition to the analysis in Figs. 11(a) and 11(b), at first the edge fluctuation is introduced by cutting a part of left-handed material in the center ring resonator on the upper edge. The corresponding edge mode is preserved nearly without any distortion, as is shown in Fig. 12(a). Secondly, similar to the case in Fig. 12(a), but both the left-handed material and the right-handed material are cut off. In this case, the selectively excited pseudo-spin-up topological one-way edge state also has not be destroyed because of the topological protection, as is shown in Fig. 12(b). At last, we consider an extreme case that the interior of the structure is hollowed out. Figure 12(c) shows the pseudo-spin-up state can propagate on the upper edge without any distortion. This robust transmission is also valid for the pseudo-spin-down state, as is shown in Fig. 12(d). Therefore, the topologically protected propagation of the edge states have been fully verified.

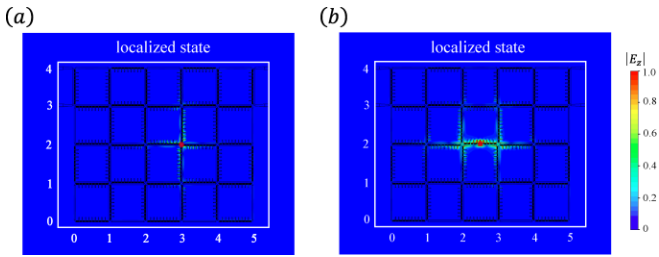

Fig. 13. Simulated electric field distributions of $|E_z|$ when the source is placed near the center of the structure at 2.5 GHz. For two different excitation positions (a, b), the wave will localize around the center of the structure. The positions of the sources are marked by the red dots.

Different from the excitation on the edge, when the source is placed near the center of the structure, the wave will localize around the exciting position, as is shown in Fig. 13. The excitation is realized by putting a monopole source mark by red dots in the bulk of the structure.

**Funding.** National Key Research Program of China (Grant No. 2016YFA0301101), National Natural Science Foundation of China (NSFC) (Grant Nos. 11674247, 61621001, 11504236, 91850206 and 11774261), Shanghai Science and Technology Committee (Grant Nos. 18ZR1442900, and 18JC1410900), the Fundamental Research Funds for the Central Universities, the Shanghai Super Postdoctoral Incentive Program, and the China Postdoctoral Science Foundation (Grant Nos. 2019TQ0232 and 2019M661605).

**Disclosures.** The authors declare no conflicts of interest.